\documentclass[a4paper, 12pt]{article}
\usepackage{amsmath, amssymb, amsfonts, amsthm, mathtools, mathrsfs, bm}
\usepackage{fullpage, relsize}
\usepackage{longtable,threeparttable,booktabs, float, rotating, multirow, adjustbox}
\usepackage{graphicx, subcaption, array}
\usepackage{tikz}
\usepackage{algorithmic}
\usepackage[linesnumbered,ruled]{algorithm2e}
\usepackage{tabularx}
\usepackage{subcaption, caption}
\usepackage{authblk}
\usepackage{setspace}
\usepackage{listings}
\usepackage{ulem}
\usepackage{url}
\usepackage{array}
\usepackage{arydshln}
\usepackage[OT1]{fontenc}

\usepackage{natbib}
\usepackage[dvipsnames]{xcolor}

\usepackage{hyperref}
\hypersetup{
colorlinks=true,
linkcolor=blue,
citecolor=blue,
urlcolor=BrickRed
}

\usepackage{todonotes} 
\usepackage{marginnote} 

\usepackage[most]{tcolorbox}

\newtheorem{definition}{Definition}[section]

\newcommand{\E}{\operatorname{E}}
\newcommand{\Var}{\operatorname{Var}}

\usepackage{titlesec}
\titleformat{\subsection}{\itshape}{\thesubsection}{0.5em}{}

\begin{document}

\title{A spatiotemporal negative binomial model with dynamic dispersion: An application to Tuberculosis infections}
\author[1]{Rodrigo B. Silva\footnote{\texttt{e-mail}: \href{mailto:rodrigo@de.ufpb.br}{rodrigo@de.ufpb.br}}}
\author[2]{Luiza S.C. Piancastelli\footnote{\texttt{e-mail}: \href{mailto:luiza.piancastelli@ucd.ie}{luiza.piancastelli@ucd.ie} (Corresponding Author). Science Centre - North, School of Mathematics and Statistics, University College Dublin, Belfield, Dublin 4, Ireland.}}
\author[2]{Wagner Barreto-Souza\footnote{\texttt{e-mail}: \href{mailto:wagner.barreto-souza@ucd.ie}{wagner.barreto-souza@ucd.ie}}}
\affil[1]{\it \normalsize Departamento de Estatística, Universidade Federal da Paraíba, Brazil}
\affil[2]{\it \normalsize School of Mathematics and Statistics, University College Dublin, Republic of Ireland}
\date{\today}
\sloppy  
\maketitle

\begin{abstract}
\noindent Tuberculosis (TB) remains a critical public health concern in Brazil, characterized by pronounced spatial heterogeneity and fluctuating temporal volatility. In this paper, we study monthly TB notifications across 61 microregions of São Paulo state from 2001 to 2024. To do this, we introduce a negative binomial spatial integer-valued generalized autoregressive conditional heteroskedastic (INGARCH) model featuring jointly dynamic conditional means and time-varying dispersion. To capture inter-regional spillovers, we incorporate both discrete adjacency structures and a novel continuous distance-based formulation leveraging the Matérn correlation function. Parameter estimation via conditional maximum likelihood employs a two-step profile-likelihood iterative scheme, demonstrating solid finite-sample performance in simulation studies. Applied to the São Paulo TB surveillance data, the framework substantially outperforms standard Poisson and fixed-dispersion spatiotemporal baselines in empirical fit and uncertainty quantification, maintaining nominal 95\% predictive coverage across both dense metropolitan centers and rural microregions. Our results reveal marked spatial heterogeneity in baseline incidence, dynamic overdispersion driven by localized outbreaks, and short-range spatial interaction decay. By accurately modeling spatiotemporal volatility, the proposed methodology provides a robust statistical tool to support public health surveillance, policy-making, and resource allocation.\\

\end{abstract}

\section{Introduction}
Tuberculosis (TB) remains one of the most persistent infectious diseases worldwide, affecting over 10 million people and causing more than 1 million deaths each year~\citep{who2025}. In Brazil, TB incidence exhibits pronounced spatial heterogeneity and complex temporal dynamics, largely driven by socioeconomic inequalities in income, sanitation, population mobility, and access to healthcare and surveillance~\citep{limaetal2024}. These features are particularly evident in the state of São Paulo, the most populous and economically developed state in the country, which accounts for a large proportion of TB cases nationwide~\citep{silvaquijano2025}. Modelling the joint spatiotemporal evolution of TB incidence is therefore essential for adequate public resource allocation and effective interventions.

From a statistical perspective, TB surveillance data present several challenges. Specifically, the number of confirmed cases is an integer-valued observation with both the mean and dispersion parameters not homogeneous in space and time; this will be justified later. This reflects TB's heterogeneous transmission dynamics, reporting variability, and episodic outbreaks. These characteristics may be further affected by structural changes in surveillance processes, such as those observed during the COVID-19 pandemic, which altered case detection and healthcare access patterns in Brazil~\citep{pontesetal2024}. In addition, temporal dependence and spatial interaction across regions are key features of the data. While frameworks for integer-valued time series and spatial modeling are independently well-established, the integration of these two domains is less developed.

To address this gap, we build upon the literature of count time series models, which has centered largely on the integer-valued generalized autoregressive conditional heteroskedastic (INGARCH) approach. Since their introduction by \citet{heinen2003} and subsequent development by \citet{ferlandetal2006}, \citet{fokianosetal2009}, and \citet{fokianostjostheim2011}, INGARCH processes have been established as a flexible framework for modelling count time series. Subsequent research has broadened this class in several directions. In the univariate setting, recent work has focused on accommodating overdispersion and heavy-tailed behaviour through flexible conditional distributions, alternative link functions, and nonlinear specifications and dynamic variance ~\citep{silvabarretosouza2019, gorgi2020, weissetal2022, weisszhu2025, barretosouzaetal2026}. In the multivariate setting, various models have been introduced to capture cross-series dependence and complex count structures~\citep{heinenrengifo2003, fokianosetal2020, jangetal2024, piancastellisilva2025}.

Despite these advances, the majority of multivariate INGARCH models do not explicitly account for spatial relationships. While they capture temporal dependence and cross-correlations, they are not designed to incorporate structured spatial interaction, a key element when data are associated with specific geographic locations. Simultaneously incorporating spatial and temporal interdependencies without a prohibitive increase in dimensionality remains an ongoing challenge. While the present paper focuses on advancing the INGARCH framework motivated by the TB incidence, several alternative approaches have also been proposed for spatio-temporal count data. For example, \citet{yangetal2025} introduced a spatio-temporal thinning-based autoregressive model to accommodate overdispersion in epidemic data, \citet{martinsetal2023} developed spatiotemporal extensions of integer-valued autoregressive moving average models, and \citet{jahn2024} integrated artificial neural networks with geographic coordinates to capture spatial heterogeneity.

Within the INGARCH framework, recent contributions have followed different strategies to capture spatial dependencies. For instance, \citet{escuderoetal2022} proposed a spatio-temporal INGARCH model where spatial dependence is modeled through a latent log-Gaussian process. Another approach involves introducing the past counts of neighboring locations directly into the rate of the process. Under this framework, if $Y_{i,t}$ denotes the counts in location $i$ at time $t$, its conditional mean $\lambda_{i,t}$ is modeled as a function of the previous counts at other locations, $Y_{j,t-1}$ for $j \neq i$. The magnitude of this spatial spillover is typically determined by a weight $w_{ij}$, which quantifies the influence of region $j$ on region $i$. In its simplest form, a binary neighbourhood structure is one where $w_{ij} = 1$ if locations $i$ and $j$ share a common boundary, and zero otherwise. Or similarly, $w_{ij} = 1/n_i$ where $n_i$ is the total number of neighbours for the target location. This neighbourhood structure has been explored, for example, in the Poisson Space-Time Autoregressive Moving-Average (PSTARMA) model by \citet{maletzetal2024}. An alternative way to specify $w_{ij}$ is as a function of the distance between locations, a continuous weighting scheme that has been successfully applied to spatial INGARCH models \citep{jahnetal2023,chenetal2023}. 

More recently, \citet{maletzetal2026} developed the \texttt{glmSTARMA} framework, which extends double generalized linear models (DGLMs) to multivariate spatio-temporal data. This approach offers considerable flexibility by supporting multiple count distributions (such as Poisson, Negative Binomial, and Binomial), flexible link functions, and dynamic dispersion. A key strength of \texttt{glmSTARMA} is its ability to estimate higher-order spatial interactions. However, spatial weight matrices have to be pre-specified for each spatial order. Another possible limitation is that \texttt{glmSTARMA} assumes spatially uniform (global) parameters drive the temporal and spatial structures. This poses difficulties for modelling count time series with large regional heterogeneity such as the Tuberculosis incidence data.

In this paper, we investigate the incidence of Tuberculosis (TB) across the state of São Paulo, Brazil, using monthly case notifications from 61 microregions between 2001 and 2024. Modeling these surveillance counts requires accommodating pronounced overdispersion, inter-regional spatial interactions, and spatiotemporally varying mean and dispersion. To capture these complex features within a unified framework, we extend the univariate time-varying dispersion INGARCH (tvd-INGARCH) model introduced by \cite{barretosouzaetal2026} to a multivariate spatiotemporal setting. Grounded in the negative binomial distribution, our proposed NB tvd-SPINGARCH model allows both the conditional mean and the dispersion parameter to evolve dynamically over time and space, resolving key limitations of conventional fixed-dispersion count models.  A central focus of our empirical investigation is determining how spatial dependencies govern disease transmission across administrative boundaries. To represent these cross-regional interactions, we formulate and compare two alternative spatial weighting structures: a discrete adjacency scheme based on shared geographic borders and a continuous, distance-dependent formulation leveraging the Matérn correlation function. By parameterizing spatial range and smoothness, the Matérn formulation allows geographic decay to be estimated directly from the data, providing a principled mechanism to test whether transmission is confined to immediate boundary spillovers or decays continuously across broader geographic ranges.  This work contributes to spatiotemporal count modeling in three main ways. First, methodologically, by expanding upon \cite{barretosouzaetal2026}, it generalizes time-varying dispersion processes to a spatial network setting, enabling joint estimation of dynamic means and dynamic variances across multiple geographic units. Second, it introduces the flexible Matérn correlation family into spatiotemporal INGARCH modeling as a continuous alternative to rigid adjacency matrices. Third, substantively, the application to São Paulo TB surveillance demonstrates that dynamic dispersion is essential for capturing localized disease volatility and obtaining reliable prediction intervals, while yielding empirical insights into spatial transmission patterns across dense urban centers and rural microregions.

The remainder of this paper is structured as follows. Section~\ref{sec:casestudy} introduces the São Paulo Tuberculosis dataset, detailing its spatial, temporal, and dispersion characteristics. Section~\ref{sec:tvd-SPINGARCH} develops the proposed NB tvd-SPINGARCH framework, outlining its spatial weighting structures, conditional maximum likelihood estimation scheme, parametric bootstrap procedure, and finite-sample simulation performance. Section~\ref{sec:application} presents the statistical analysis of the monthly TB notifications, evaluating the performance of alternative spatial formulations against existing spatiotemporal count benchmarks. Finally, Section~\ref{sec:conclusion} offers concluding remarks.

\section{Case study}\label{sec:casestudy}

The TB data were obtained from the Department of Information and Informatics of the Brazilian Unified Health System (DATASUS), which is the official repository for healthcare information in the country. For TB, cases are reported on a monthly basis and per microregion of federal states. We focus on the state of São Paulo and the period from January 2001 to December 2024 (288 time points). Throughout this work, we will denote by $Y_{it}$ the TB cases recorded in spatial unit $i$ at time $t$. Further, $\mathbf Y_t = (Y_{1t}, \ldots, Y_{pt})^\top$ is the vector of counts observed for $t=1, \ldots,T$, $p=61$, and $T = 288$. 

\begin{figure}[ht!]
    \centering
    \includegraphics[width=0.91\linewidth]{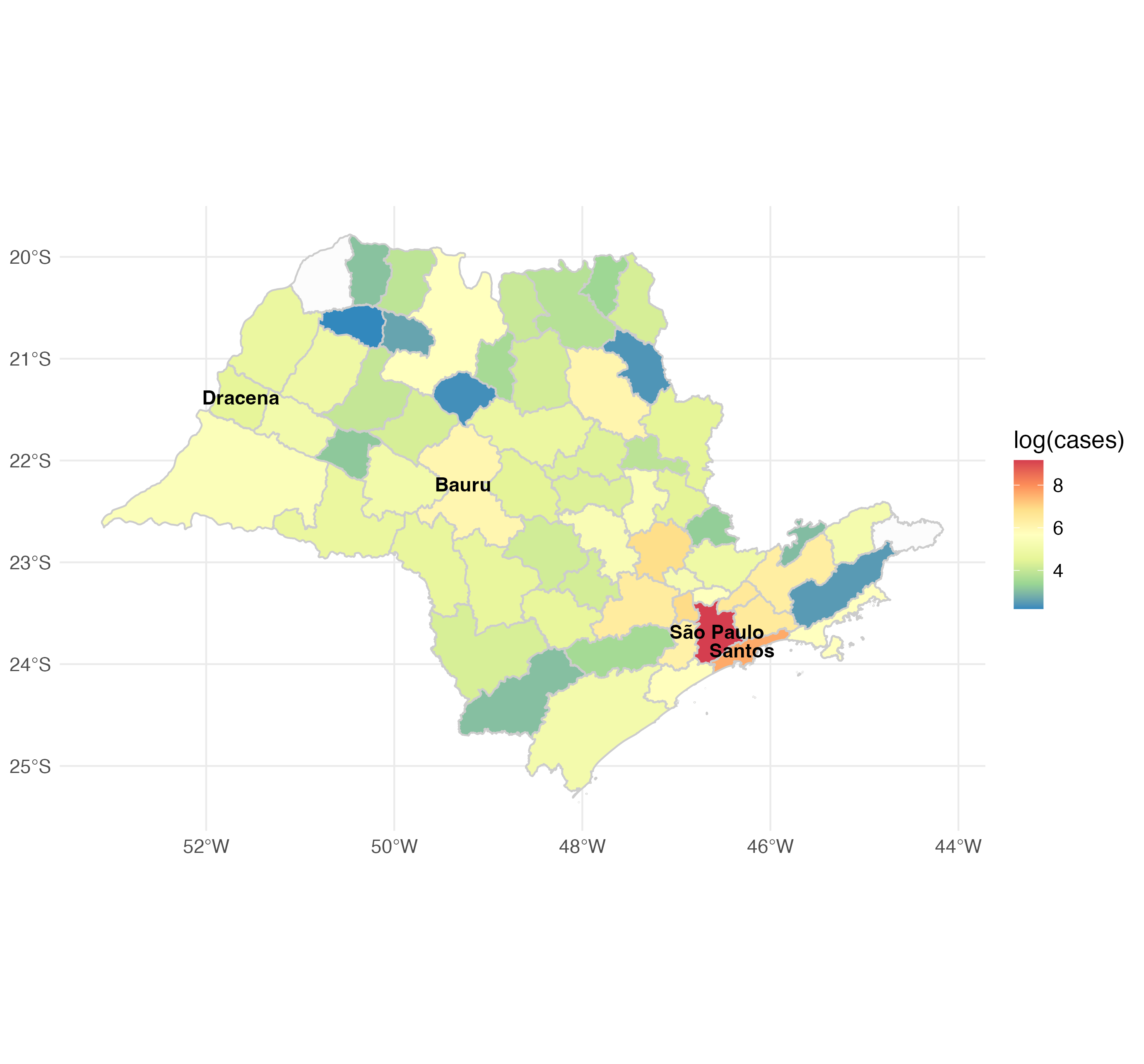}
    \caption{\it Tuberculosis cases in microregions of the state of São Paulo in 2024. }
    \label{fig:mapsp2024}
\end{figure}

Figure~\ref{fig:mapsp2024} illustrates the spatial distribution of the total number of tuberculosis cases registered in 2024 across 61 microregions of the state of São Paulo. As expected, counts concentrate in dense urban centers, particularly the metropolitan region of the state's capital city (São Paulo city). We highlight that two of the 63 total São Paulo microregions were excluded from the analysis due to data anomalies and inconsistencies.

\begin{figure}[ht!]
\centering
\includegraphics[scale=0.7]{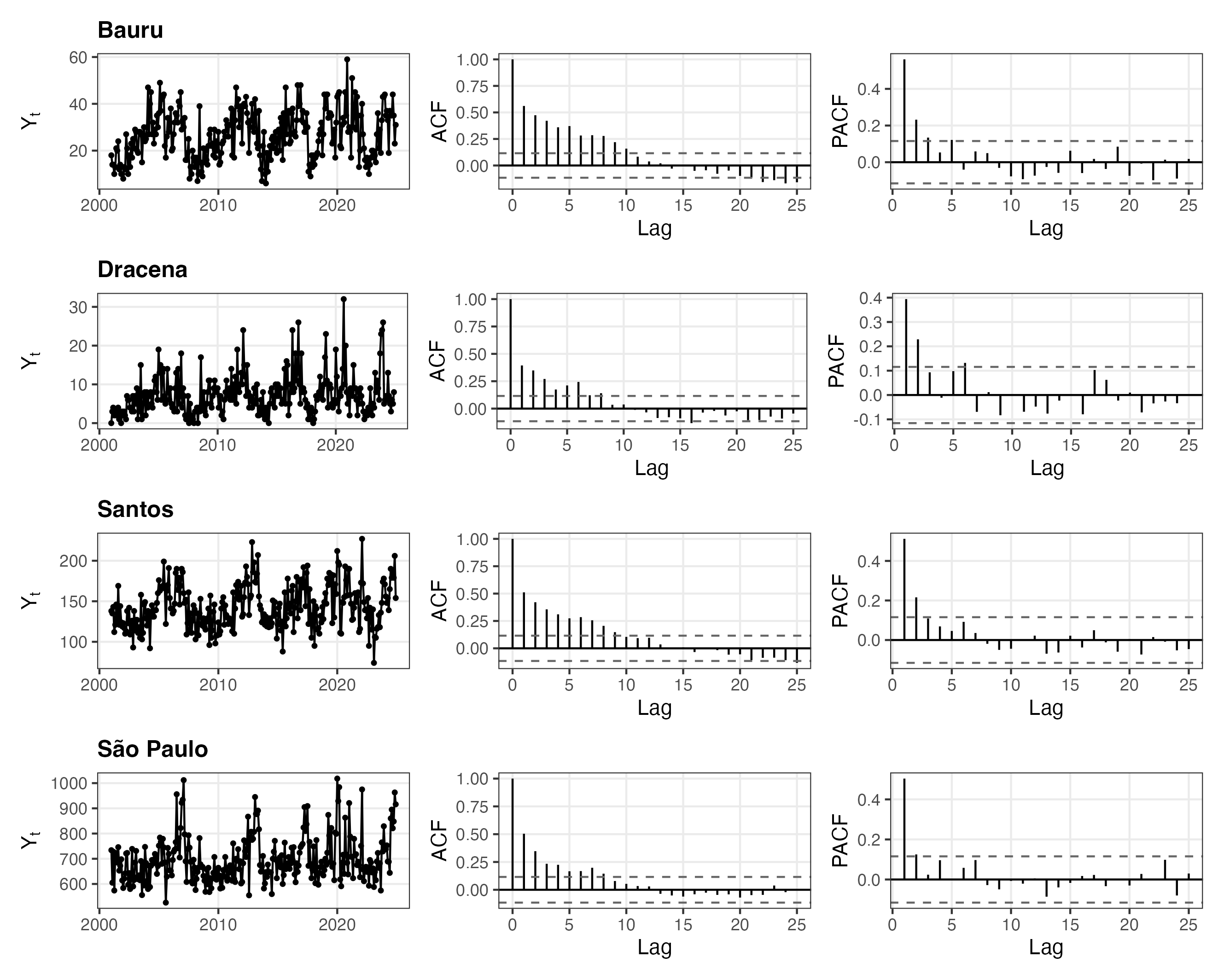}
\caption{\it Monthly tuberculosis counts for Bauru, Dracena, Santos and São Paulo, with corresponding ACFs and PACFs.}
\label{fig:saopaulo}
\end{figure}

\begin{figure}[ht!]
\centering
\includegraphics[scale=0.35]{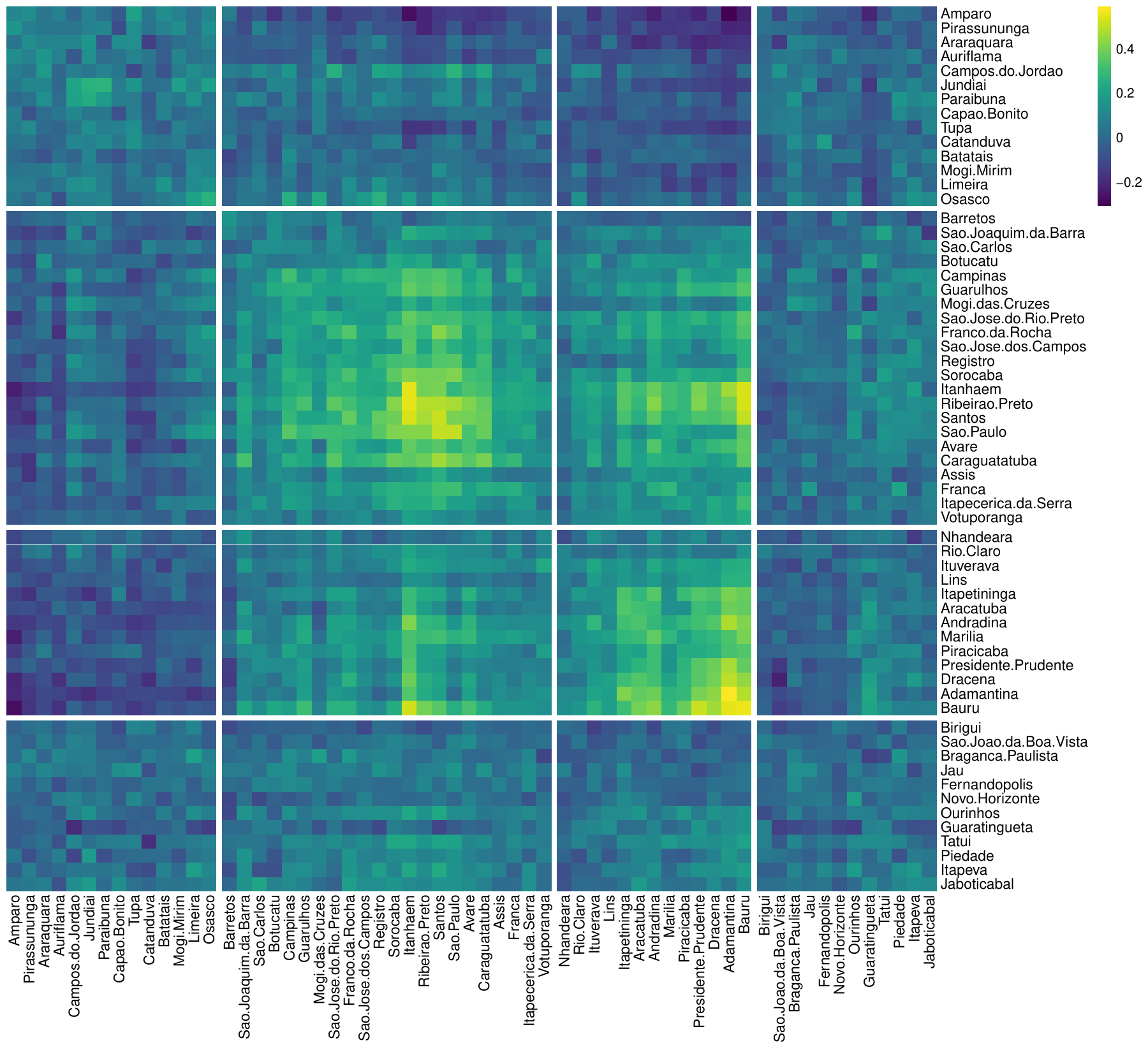}
\caption{\it Heatmap of the lag-1 cross-correlation matrix between the observed count series across the microregions. Colour intensity indicates the strength and direction of the correlations.}
\label{fig:SP_heatmap}
\end{figure}

Figure~\ref{fig:saopaulo} presents the plot of the monthly tuberculosis cases (first column) alongside their ACFs and PACFs (second and third columns) for distinct metropolitan, coastal, and inland microregions. Crucially, temporal dependence extends beyond individual series. Figure~\ref{fig:SP_heatmap} explores cross-region dynamics using lag-one cross-correlations, $\mathrm{Cor}(Y_{it}, Y_{j,t-1})$ for $i \neq j$, to assess whether past counts in region $j$ inform current cases in region $i$. Correlations are consistently positive and strongest among neighboring pairs, highlighting the predictive utility of spatial lag variables.

A closer examination of monthly counts across the selected microregions is provided in Table~\ref{tab:summary_stats}. Beyond reflecting population scale, the index of dispersion (ID) suggests region-specific dispersion behaviours. Although every representative microregion displays overdispersion ($\mathrm{ID} > 1$), the magnitude ranges considerably, from $3.65$ in Dracena to $11.61$ in São Paulo.

\begin{figure}[ht!]
\centering
\includegraphics[scale=0.41]{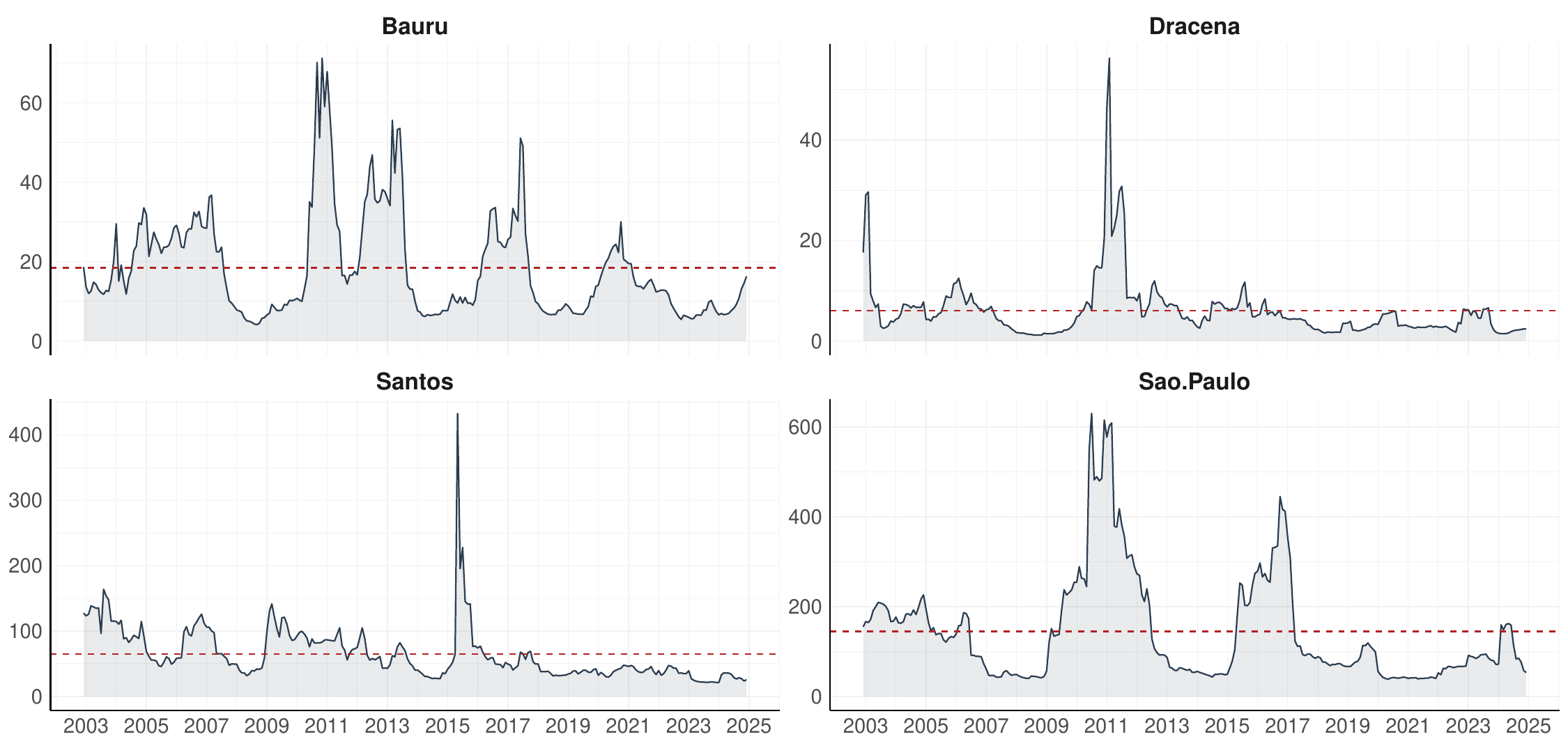}
\caption{\it Empirical dispersion index of monthly TB counts for four representative microregions in the state of São Paulo: Bauru, Dracena, Santos and São Paulo. The horizontal red line represents the average dispersion index for each microregion over the study period.}
\label{fig:dispersion_index}
\end{figure}

In addition to spatial heterogeneity, dispersion also exhibits temporal dynamics, as shown in Figure~\ref{fig:dispersion_index}. This figure displays estimates of the dispersion parameter of a Negative Binomial distribution computed over 24-month moving windows. The results suggest that the dispersion varies over time, differing in both magnitude and volatility across locations. This parameter is inversely related to overdispersion; thus, peaks correspond to periods of lower relative variance, whereas troughs indicate periods of heightened overdispersion.

\begin{table}[ht!]
\centering
\small
\renewcommand{\arraystretch}{1.2}
\begin{tabular}{lrrrr}
\toprule
\textbf{Microregion} & \textbf{Mean} & \textbf{Variance} & \textbf{Range} & \textbf{ID} \\ 
\midrule
São Paulo & 698.50 & 8110.45 & 526--1018 & 11.61 \\
Santos    & 143.40 &  721.02 &  74--227  &  5.03 \\
Bauru     &  26.87 &  107.54 &   6--59   &  4.00 \\
Dracena   &   7.27 &   26.53 &   0--32   &  3.65 \\ 
\bottomrule
\end{tabular}
\caption{Summary statistics of monthly tuberculosis counts for representative microregions in São Paulo (2001--2024); ID denotes the index of dispersion.}
\label{tab:summary_stats}
\end{table}

Overall, this empirical analysis highlights key features of the TB data. First, temporal and spatial interdependencies indicate that $Y_{it}$ should be modeled as a function of both its own past values as well as historical counts from other locations. How and which locations to consider is a question that we explore in this paper. Second, TB counts are overdispersed, and the dispersion behaviour is both location-specific and dynamic over time. Integrating these aspects into a single modeling framework motivates the proposed methodology.


\section{Spatiotemporal negative binomial model with dynamic dispersion}\label{sec:tvd-SPINGARCH}
\vspace{-.2cm}

The proposed model is introduced in this section. We start with preliminary concepts, recalling the Negative-Binomial (NB) INGARCH model and then introducing the time-varying dispersion (tvd) INGARCH framework of \cite{barretosouzaetal2026}, which will be extended here to accommodate spatial dependence. 

A random variable $Y$ follows a negative binomial (NB) distribution with respective mean and dispersion parameters $\mu > 0$ and $\phi > 0$ if its probability mass function (pmf) assumes the form 
\begin{equation}\label{eq:nb}
    P(Y=y) = \frac{\Gamma(y+\phi)}{y!\Gamma(\phi)}\left(\frac{\mu}{\mu + \phi}\right)^y \left(\frac{\phi}{\mu + \phi}\right)^\phi, \quad y \in \mathbb{N}_0,
\end{equation}
and we denote $Y \sim \mathrm{NB}(\mu, \phi)$.
Under the parameterization in~\eqref{eq:nb}, we have that
\begin{equation*}\label{eq:nbmoments}
    \E(Y) = \mu \quad \text{and} \quad \Var(Y) = \mu\left(1+\frac\mu{\phi}\right),
\end{equation*}
so the distribution is overdispersed ($\Var(Y)>\E(Y)$) and approaches the Poisson case (equidispersed, $\Var(Y) = \E(Y)$) as $\phi \rightarrow \infty$. 

Let $\{Y_t\}_{t\geq1}$ be a count time series and $\mathcal{F}_{t-1}$ be the $\sigma$-algebra generated by $Y_{t-1},\ldots,Y_1$, for $t\geq2$. Then, $\{Y_t\}_{t\geq1}$ follows an (univariate) NB-INGARCH model if $Y_t \mid \mathcal{F}_{t-1} \sim \mathrm{NB}(\mu_t, \phi)$, for $t\geq2$, where the conditional mean $\mu_t$ evolves according to the INGARCH recursion
\begin{equation} \label{eq:mu_t}
    \mu_t = \omega + \sum_{i=1}^p \alpha_i Y_{t-i} + \sum_{j=1}^q \beta_j \mu_{t-j},
\end{equation}
where the parameters $\omega, \alpha_i, \beta_j > 0 \quad \forall i,j$ are the intercept, autoregressive, and feedback effects, respectively.
Alternatively, the INGARCH model can be defined through a log-linear form, a specification that removes parameter constraints and eases the inclusion of covariates. In this case, 
\begin{equation} \label{eq:mu_t_log}
    \log\mu_t = \omega + \sum_{i=1}^p \alpha_i \log(Y_{t-i} + 1) + \sum_{j=1}^q \beta_j \log\mu_{t-j},
\end{equation}
which is known as the log-linear INGARCH model; for instance, see \cite{fokianostjostheim2011}. 

Although widely applied across count time series analysis, standard INGARCH models assume a static dispersion, i.e. $\phi$ is fixed and does not change in time. This may not be realistic for real-world data, where both the mean and dispersion may exhibit temporal fluctuations. The recent extension by \citet{barretosouzaetal2026}, the \textit{time-varying dispersion INGARCH} (tvd-INGARCH) model, addresses this with an additional log-linear recursion for $\phi$, which is given by
\begin{equation*}\label{eq:phi_t}
    \log\phi_t = \delta_0 + \sum_{i=1}^r \delta_i \log(Y_{t-i}+1) + \sum_{j=1}^s \gamma_j \log\phi_{t-j},
\end{equation*}
similarly to (\ref{eq:mu_t_log}).

Building upon this, our proposed framework assumes that, conditional on the historical information $\mathcal{F}_{t-1}$, the count time series $Y_{1t}, \ldots, Y_{pt}$ are independent negative binomial processes. While the spatial structure in the conditional mean draws inspiration from \cite{jahnetal2023} and \cite{maletzetal2026}, our approach involves novel key ingredients. First, it introduces a dynamic spatiotemporal recursion for the dispersion parameter to account for evolving volatility across locations. Second, it departs from rigid spatial definitions by introducing alternative, parametric distance-based weighting schemes alongside discrete adjacency structures to model cross-regional transmission. The proposed new spatiotemporal count model is formally defined in what follows.

\begin{definition}[{\it tvd-SPINGARCH process}]\label{def:tvd-SPINGARCH}
    Let $\{\mathbf{Y}_t\}_{t \geq 1} = \{(Y_{1t}, \ldots, Y_{pt})^\top\}_{t \geq 1}$ be a $p$-dimensional count process and define the $\sigma$-algebra $\mathcal{F}_{t-1} = \sigma(\mathbf Y_{t-1}, \ldots, \mathbf Y_1)$, for $t\geq2$. We define the time-varying dispersion spatial INGARCH (tvd-SPINGARCH) process by assuming conditional independence of $Y_{1t}, \ldots, Y_{pt}$ given $\mathcal{F}_{t-1}$ and
    \begin{equation}\label{eq:tvdSPINGARCH}
	\begin{cases}
	 \,\,Y_{it}|\mathcal{F}_{t-1} \sim \mbox{NB}(\mu_t, \phi_t),\,\, \mu_{it} = \exp(\lambda_{it}),\,\, \phi_{it} = \exp(\nu_{it}),\\
     \,\, \lambda_{it} = d_i^{(\lambda)} + a_{i}^{(\lambda)} \log(Y_{i,t-1}+1) + b_i^{(\lambda)}\sum_{j\neq i} w_{ij} \log(Y_{j,t-1} + 1),\\
     \,\, \nu_{it} = d_i^{(\nu)} + a_i^{(\nu)} \nu_{i,t-1} + b_i^{(\nu)} \sum_{j\neq i} w_{ij} \log(Y_{j,t-1} + 1),
    \end{cases}
\end{equation}
for $t\geq1$ and $i=1,\ldots,p$, where the $w_{ij}$'s are appropriately chosen weights. In this formulation, $d_i^{(\cdot)}$ denote region-specific intercepts. The coefficients $a_i^{(\cdot)}$ define the autoregressive component of the latent processes $\{\lambda_{it}\}$ and $\{\nu_{it}\}$, while $b_i^{(\cdot)}$ govern the spatial interaction effects via weighted contributions from neighbouring regions, given by $\sum_{j \ne i} w_{ij} \log(Y_{j,t-1} + 1)$.
\end{definition}

Specific formulations for constructing the spatial weights $w_{ij}$ are detailed in the following subsection.

\subsection{Spatial dependence structure}\label{sec:spatialdependence}

The spatial weights $w_{ij}$ quantify the strength of transmission from unit $j$ at time $t-1$ to unit $i$ at time $t$. As mentioned previously, a simple approach is to set $w_{ij} = 1/n_i$ if units $i$ and $j$ are adjacent, and $w_{ij} = 0$ otherwise, where $n_i$ denotes the number of neighbours of unit $i$. Normalisation by $n_i$ is done so that weights are normalised, $\sum_{i\neq j} w_{ij} = 1$, see~\cite{pauletal2008}. Other schemes include the power-law formulation proposed by~\cite{meyerheld2014}, where weights are defined based on the geographical distance between $i$ and $j$, $o_{ij}$, and $w_{ij}$ is proportional to $o_{ij}^{-\delta}$. Another option that uses $o_{ij}$ but normalises weights is $w_{ij} = (1+o_{ij})^{-\delta}/ \sum_{k=1}^p (1+o_{ik})^{-\delta}$ ~\citep{bracherheld2022}.

We propose two forms of defining $w_{ij}$ for the TB data.
The first is the classical $w_{ij} = 1/n_i$ stipulation, where microregions are considered neighbours if they share borders. To generalize this structure to point coordinates without predefined administrative boundaries, we include the option of constructing a geometrically defined adjacency system using Voronoi diagrams as done by \citet{piancastellietal2024}. Given a set of spatial centroids $\{\mathbf{s}_1, \ldots, \mathbf{s}_n\}$, the spatial domain is partitioned into $n$ non-overlapping regions $\mathcal{V}_i$:
\begin{equation}\label{eq:voronoi}
    \mathcal{V}_i = \left\{ \mathbf{s} \in \mathbb{R}^2 : \|\mathbf{s} - \mathbf{s}_i\| \leq \|\mathbf{s} - \mathbf{s}_j\|, \; \forall j \neq i \right\}.
\end{equation}
Under this partition, two locations are considered neighbours if their corresponding Voronoi polygons $\mathcal{V}_i$ and $\mathcal{V}_j$ share a boundary. 

The second proposed approach defines the spatial weights in terms of $o_{ij}$ using the Matérn correlation form. The Matérn function is widely applied in spatial statistics \citep{apanasovichetal2012,bevilacquaetal2022,wangetal2023,porcuetal2024}, often setting the form of the spatial covariance matrix for models based on the Gaussian process. However, its application as a weighting mechanism has not been explored to the best of our knowledge. The Matérn correlation function takes the form
\begin{equation}\label{eq:materncor}
w_{ij} = \mathcal{M}_{\alpha, \beta}(o_{ij}) = \frac{2^{1-\alpha}}{\Gamma(\alpha)} \left(\sqrt{2\alpha}\frac{o_{ij}}{\beta}\right)^{\alpha} \mathcal{K}_{\alpha}\left(\sqrt{2\alpha}\frac{o_{ij}}{\beta}\right),
\end{equation}
where $\alpha>0$ and $\beta>0$ control the smoothness and scale of the Matérn function, respectively, and $\mathcal K_\rho(z) = \dfrac{1}{2}\displaystyle\int_0^\infty u^{\rho-1}\exp\{-z(u + u^{-1})/2\}du$ is the modified Bessel function of the second kind, where $\rho\in\mathbb R$ and $z>0$. The Matérn form for $w_{ij}$ follows from Equation (\ref{eq:materncor}) and row-wise normalisation. That is, let $\widetilde{w}_{ij} = \mathcal{M}_{\alpha,\beta}(o_{ij})$ for $i \neq j$ with $\widetilde{w}_{ii}=0$. Then, $w_{ij} = \widetilde{w}_{ij}/\sum_{k \neq i} \widetilde{w}_{ik}$.

\begin{figure}[ht!]
\centering
\includegraphics[scale=0.47]{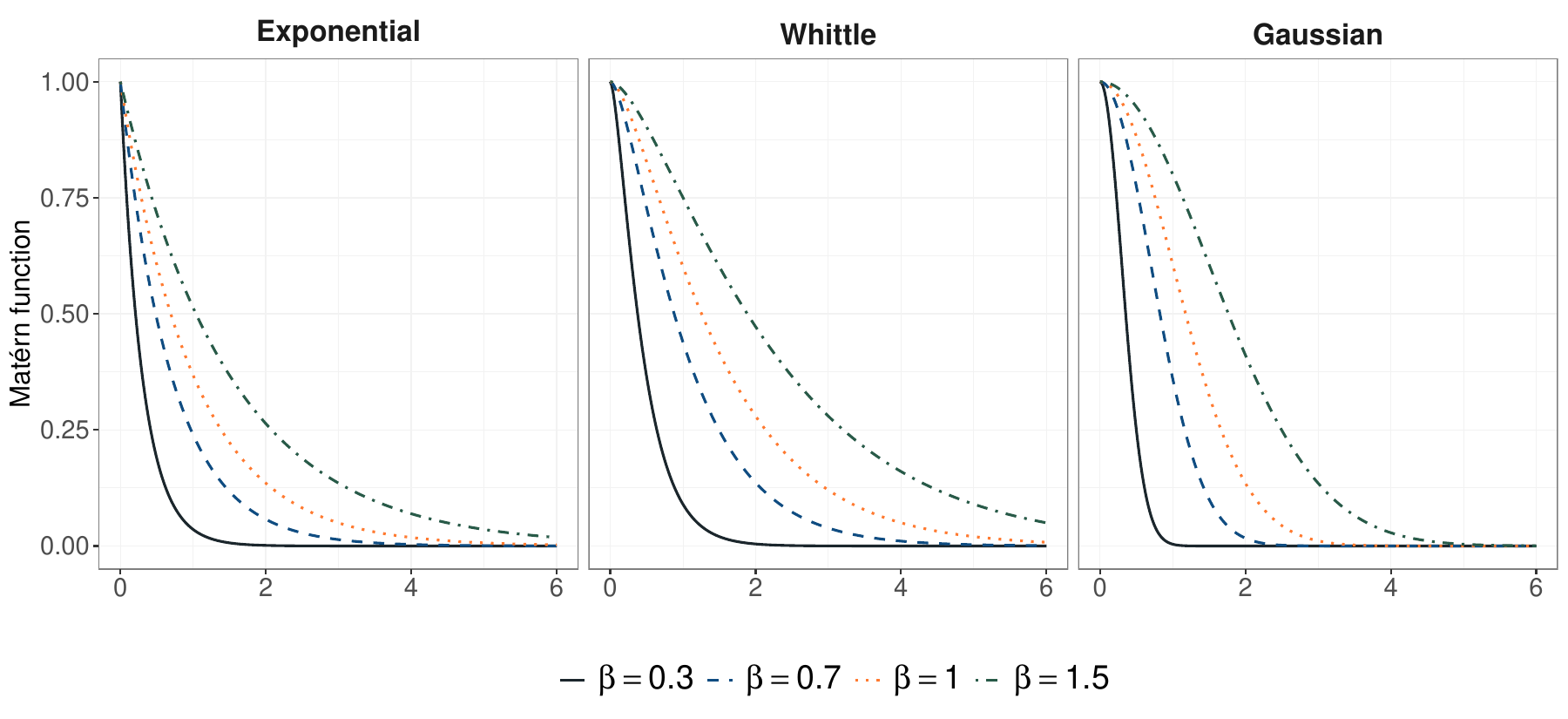}
\caption{\it Plots of particular cases of the Matérn function for some values of $\beta$, namely exponential $(\alpha=0.5)$, Whittle $(\alpha=1)$ and Gaussian $(\alpha \rightarrow \infty)$.}
\label{fig:matern_cases}
\end{figure}

Adopting a parametric form for $w_{ij}$ enables formal statistical inference on the mechanisms governing spatial spread. Here, the flexible Matérn function explicitly models how spatial correlation decays with distance (governed by $\alpha$) and determines the overall range of spatial dependence (controlled by $\beta$). Moreover, it encompasses several well-known correlation structures as special cases, including the exponential ($\alpha = 1/2$), Whittle ($\alpha = 1$), and Gaussian ($\alpha \rightarrow \infty$) covariances, whose behaviours are illustrated in Figure~\ref{fig:matern_cases} for some values of $\beta$.  

Naturally, the Matérn form for $w_{ij}$ introduces additional challenges for statistical inference. In contrast with the adjacency-based approach, weights are now a function of unknown parameters. This aspect is carefully handled in the next subsection.


\subsection{Statistical inference}

Consider $\{\mathbf{Y}_t\}_{t = 1}^T = \{(Y_{1t}, \dots, Y_{pt})^\top\}_{t = 1}^T$ a sample of size $T$ observed from a tvd-SPINGARCH process, with realizations $\mathbf{y}_t = (y_{1t}, \dots, y_{pt})^\top$. The conditional log-likelihood function is given by
\begin{align*}\label{eq:loglik} \nonumber
    \ell(\bm\theta) &\equiv \sum_{t=2}^T \sum_{i=1}^p \log P(Y_{it} = y_{it} \mid \mathcal{F}_{t-1}) \\[0.2em] 
    &\propto \sum_{t=2}^T \sum_{i=1}^p \left\{ y_{it}\log\left(\frac{\mu_{it}}{\mu_{it} + \phi_{it}}\right) + \phi_{it} \log\left(\frac{\phi_{it}}{\mu_{it} + \phi_{it}}\right) + \log\Gamma(y_{it} + \phi_{it}) - \log\Gamma(\phi_{it}) \right\},
\end{align*}
where the parameter vector is $\bm\theta = (\bm\vartheta^\top, \alpha, \beta)^\top$ with $\bm\vartheta = (\mathbf{d}^{(\lambda)}, \mathbf{a}^{(\lambda)}, \mathbf{b}^{(\lambda)}, \mathbf{d}^{(\nu)}, \mathbf{a}^{(\nu)}, \mathbf{b}^{(\nu)})^\top$ for the Mat\'{e}rn formulation, and reduces to $\bm\theta = \bm\vartheta$ for the adjacency version. 

A key computational distinction between the estimation procedures of the two versions of the model arises from the fact that the spatial weight matrix $\mathbf{W}$ is fixed and parameter-free under the adjacency model. This implies that the log-likelihood function decomposes into a sum of $p$ independent regional contributions, $\ell(\bm\theta) = \sum_{i=1}^p \ell_i(\bm\theta_i)$ where $\bm\theta_i = (d_i^{(\lambda)}, a_i^{(\lambda)}, b_i^{(\lambda)}, d_i^{(\nu)}, a_i^{(\nu)}, b_i^{(\nu)})^\top$. Because no parameters are shared across distinct locations ($i \neq j$), maximum likelihood estimation reduces to $p$ isolated, low-dimensional optimization tasks that can be efficiently solved via parallel or block-wise optimization.

In contrast, the Mat\'{e}rn formulation embeds the spatial range and smoothness parameters $(\alpha, \beta)$ directly into $\mathbf{W}(\alpha, \beta)$. Because $(\alpha, \beta)$ are shared across all spatial units, they couple the dynamic conditional means $\mu_{it}$ and dispersion terms $\phi_{it}$ network-wide. Consequently, the log-likelihood function does not factorize. 

In practice, joint optimization of the full log-likelihood $\ell(\bm\theta)$ with respect to $(\alpha, \beta)$ and $\bm\vartheta$ presents severe numerical challenges. First, the optimization scale is substantial, as $\bm\vartheta$ alone contains $6p$ site-specific coefficients. Second, strong coupling between $(\alpha, \beta)$ and $\bm\vartheta$ creates an ill-conditioned likelihood surface, making unconstrained joint optimization unstable.

To overcome this, we adopt a profile-likelihood optimization strategy. This approach alternates between fixing $(\alpha, \beta)$ to optimize $\bm\vartheta$ and fixing $\bm\vartheta$ to optimize $(\alpha, \beta)$. In addition to breaking coupling, parallelized site-level sub-problems are possible as in the adjacency weighting scheme. The algorithm iterates this alternating scheme until convergence, transforming a high-dimensional joint estimation problem into a sequence of stable, low-dimensional tasks.

The complete estimation routine for the Mat\'{e}rn tvd-SPINGARCH model is summarized in Algorithm~\ref{alg:estimation}. In Step 1, the spatial range and smoothness parameters $(\alpha, \beta)$ are updated by maximizing the full log-likelihood conditional on the current estimates of $\bm\vartheta$. In Step 2, given the updated spatial weight structure, the regional parameters $\bm\vartheta_i$ are re-estimated independently across locations $i = 1, \dots, p$. The algorithm alternates between Steps 1 and 2 until the relative log-likelihood change falls below a preset tolerance $\varepsilon>0$.

\begin{tcolorbox}[
 float=htp,
  colback=gray!3,
  colframe=gray!80,
  boxrule=0.2pt,
  arc=2pt
]
\begin{algorithm}[H]
\small
\caption{Iterative estimation of the tvd-SPINGARCH model parameters}
\label{alg:estimation}
\DontPrintSemicolon
\SetAlgoLined
\LinesNotNumbered
\KwIn{Observed spatiotemporal counts $\{\mathbf Y_t\}_{t=1}^T$, distance matrix $\mathbf{D}$, maximum number of iterations $K$, tolerance $\varepsilon$.}
\vskip1mm
1. (\textbf{Initial guess}): Input starting values $\boldsymbol{\theta}^{(0)}$.\;
\vskip1mm
2. (\textbf{Iterative parameter estimation}): Initialize log-likelihood $\ell_0 = -\infty$.
\vskip1mm
\For{$\kappa \gets 1$ \KwTo $K$}{

  \textbf{(Step 1)}: Update spatial parameters:\;
  \hskip6mm Maximize the log-likelihood with respect to $(\alpha, \beta)$ given $\boldsymbol{\vartheta}^{(\kappa-1)}$.\;
  \hskip6mm Obtain updated $(\alpha^{(\kappa)}, \beta^{(\kappa)})$.\;
\vskip1mm
  \textbf{(Step 2)}: Update $\boldsymbol{\vartheta}$ for each unit:\;
  \For{$i \leftarrow 1$ \KwTo $p$}{
    Maximise the log-likelihood with respect to $\boldsymbol{\vartheta}_i$
    given current $(\alpha^{(\kappa)}, \beta^{(\kappa)})$.\;
    Obtain updated $\boldsymbol{\vartheta}_i^{(\kappa)}$.
  }
\vskip1mm
  \textbf{(Step 3)}: Compute full log-likelihood $\ell_\kappa$ using updated parameters:\;
  \hskip6mm Compute $\ell_\kappa\equiv \ell(\boldsymbol{\theta}^{(\kappa)})$.
\vskip1mm 
  \textbf{(Step 4)}: Convergence check:\;
  \hskip6mm Repeat Steps 1 to 3 until some pre-specified stopping criterion, for instance,\\ \hskip6mm $|\ell_\kappa - \ell_{\kappa-1}| / (|\ell_{\kappa-1}| + 10^{-10}) < \varepsilon$, is satisfied, or $\kappa = K$. 
}
\KwOut{$\boldsymbol{\theta}^{(\kappa)}.$}
\end{algorithm}
\end{tcolorbox}

Another aspect that deserves careful consideration is parameter initialization. In particular, we observed that the Mat\'{e}rn parameters $(\alpha, \beta)$ are susceptible to flat log-likelihood regions in which $\bm\vartheta$ overwhelms the spatial signal. To mitigate this, we construct a principled grid of candidate values for $(\alpha, \beta)$ initialisation. Normalizing the distance matrix to $[0,1]$ ensures that these parameters operate on a unit scale, facilitating the grid choice. Specifically, the definition $\alpha \in \{0.3, 0.5, 0.7, 1.0, 1.3\}$ and $\beta \in \{0.03, 0.05, 0.08, 0.12, 0.20\}$ covers smoothness levels and range values spanning from very localized to moderately long-range spatial dependencies. For each grid pair, the corresponding regional parameters $\bm\vartheta$ are conditionally estimated. The configuration achieving the maximum global log-likelihood and its $\bm\vartheta$ estimates is selected as the starting vector for $\bm\theta^{(0)}$. This grid-based pre-search substantially improves parameter identifiability, avoids flat regions of the objective surface, and stabilizes the subsequent optimization.

Finally, uncertainty associated with the final point estimates is derived via parametric bootstrap. Let $\widehat{\boldsymbol\theta}$ denote the estimates obtained from fitting Algorithm \ref{alg:estimation} to the observed data. Replications of data are generated by simulating tvd-SPINGARCH trajectories at $\widehat{\boldsymbol\theta}$. This follows the schematic detailed in Algorithm \ref{alg:datageneration}. For each replicate $b = 1, \ldots, B$, we re-estimate the parameter vector, obtaining $\widehat{\boldsymbol\theta}^{(b)}$. The empirical covariance matrix of $\{\widehat{\boldsymbol\theta}^{(b)}\}_{b=1}^B$ is then used as an estimate of the sampling variability of $\widehat{\boldsymbol\theta}$, and standard errors are computed as the square roots of the diagonal elements of this matrix.
Similar bootstrap-based strategies for quantifying uncertainty in time series models have been widely used in the literature, particularly in settings involving dependent data, nonlinear dynamics, and complex model structures; see, e.g., \citet{maiaetal2021} and \citet{barretosouzaetal2023}.

\begin{tcolorbox}[
  float=ht,
  colback=gray!3,
  colframe=gray!50,
  boxrule=0.2pt,
  arc=2pt
]
\begin{algorithm}[H]
\small
\caption{Simulating a trajectory of length $T$ and dimension $p$ from the NB tvd-SPINGARCH process for selected parameter values.}
\label{alg:datageneration}
\vskip1mm
\DontPrintSemicolon
\SetAlgoLined
\SetNoFillComment
\LinesNotNumbered
\KwIn{
Trajectory length $T$; \\
\vskip1mm
\quad Dynamic processes parameters: $\{d_i^{(\lambda)}, a_i^{(\lambda)}, b_i^{(\lambda)}\}_{i=1}^p$ and $\{d_i^{(\nu)}, a_i^{(\nu)}, b_i^{(\nu)}\}_{i=1}^p$;\\
\vskip1mm
\quad Spatial parameters: $\alpha > 0$, $\beta > 0$; \quad \tcp{\small \textcolor{Black}{if Matérn function.}}
}
\vskip3mm
1. Initialize $\lambda_{i1}$ and $\nu_{i1}$, for $i = 1,\ldots,p$. Set $\mu_{i1} \gets \exp(\lambda_{i1})$, $\phi_{i1} \gets \exp(\nu_{i1})$;
\vskip1mm
2. Simulate $Y_{i1} \sim \text{NB}(\mu_{i1}, \phi_{i1})$, for $i = 1,\ldots,p$;
\vskip1mm
3. Compute $\mathbf{W}$ from Voronoi diagrams~\eqref{eq:voronoi} or Matérn function~\eqref{eq:materncor};
\vskip1mm
\For{$t \gets 2$ \KwTo $T$}{
\For{$i \gets 1$ \KwTo $p$}{
Update $\lambda_{it}, \nu_{it}, \mu_{it}, \phi_{it}$ according to the framework in~\eqref{eq:tvdSPINGARCH};
\vskip1mm
Simulate $Y_{it} \sim \text{NB}(\mu_{it}, \phi_{it})$;
}}
\vskip1mm
\KwOut{Simulated count series $\{Y_{it}\}$, for $i=1,\ldots,p$ and $ t=1,\ldots,T$;}
\end{algorithm}
\end{tcolorbox}


\subsection{Simulation study}\label{subsec:simulation}

This section investigates the finite-sample performance of the proposed estimation procedure via a Monte Carlo simulation. Our simulation studies explore the Matérn and adjacency-based spatial weighting constructions in Settings I and II, respectively. For each scenario, $N = 500$ Monte Carlo replications are employed. All numerical experiments and statistical analyses reported in this paper were conducted using the \texttt{R} programming language, version 4.5.2.

The number of spatial units in this study is set to $p = 20$, and sample sizes $T= 500,1000$ are considered. To simplify the interpretation and visualisation of the results, identical theoretical values were assigned to INGARCH parameters across $i$. In Setting I, $d^{(\lambda)} = 0.05, a^{(\lambda)}=0.2, b^{(\lambda)}= 0.1, d^{(\nu)}=0,  a^{(\nu)}=0.1,  b^{(\nu)}=0.1, \alpha = 1.4, \beta = 0.47$, and in Setting II, $d^{(\lambda)} = 0.4, a^{(\lambda)}=-0.2, b^{(\lambda)}= 0.1, d^{(\nu)}=-0.3,  a^{(\nu)}=0.1,  b^{(\nu)}=0.5$.

\begin{figure}[ht!]
\centering
\includegraphics[width=0.9\linewidth]{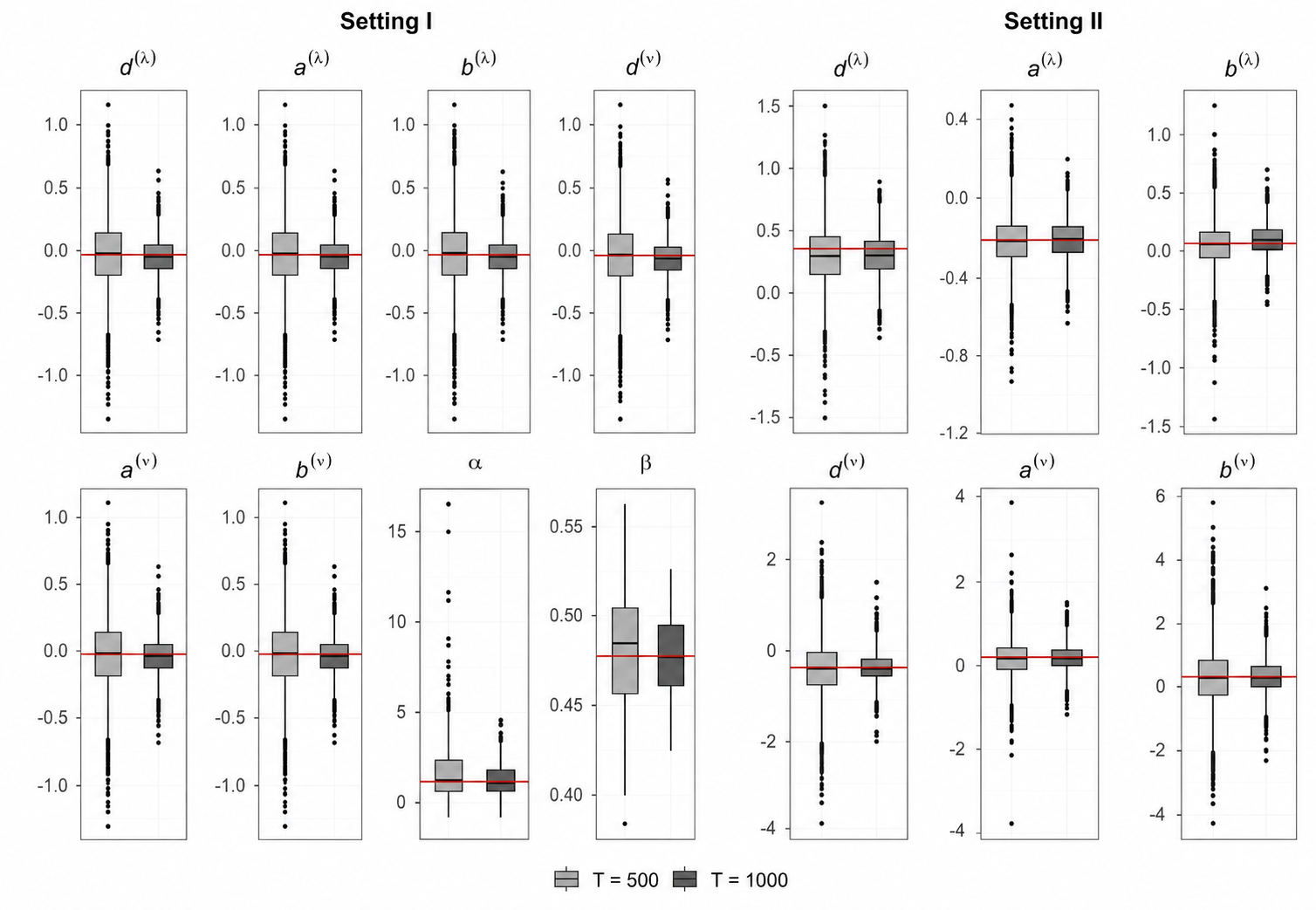}
\caption{\it Boxplots of parameter estimates for Scenarios I (Matérn-based weights) and II with $T = 500, 1000$.
Horizontal lines represent the true parameter values.}
\label{fig:boxplots_c1c2}
\end{figure}

Figure \ref{fig:boxplots_c1c2} illustrates the results. The reported values correspond to the average, over replications, of the component-wise means obtained across the 20 spatial locations. We observe that, despite the large number of parameters, the CML estimator performs well for the dynamic process parameters $\lambda_t$ and $\nu_t$, with empirical means close to the true parameter values and the dispersion of the estimates decreasing as the sample size increases. For the Matérn-based scenario, the spatial parameters exhibit greater variability. However, as the sample size increases, its empirical mean approaches the true value while its standard deviation decreases. Overall, the results confirm that the estimation procedure is adequate for both specifications.

\section{Analysis of São Paulo Tuberculosis data}\label{sec:application}

We now apply the proposed NB tvd-STINGARCH framework to the monthly Tuberculosis (TB) counts introduced in Section~\ref{sec:casestudy}, directly addressing the spatiotemporal features and dynamic variability that motivated its development.
To benchmark its performance, we compare it to two key competitors. The first is the \texttt{glmSTARMA} framework \citep{maletzetal2026}, which also allows for time-varying dispersion and spatial effects. Despite these similarities, there are key distinctions that separate it from the tvd-SPINGARCH. These will be made explicit shortly. As a benchmark, we also estimate a Poisson SPINGARCH model with adjacency-based spatial weights. The latter is connected to the work by \cite{maletzetal2024}.

The \texttt{glmSTARMA} model is fitted using its dedicated \textsf{R} package~\citep{maletzetal2026}. To ensure direct comparability with the tvd-SPINGARCH, we specify a Negative Binomial distribution family with dynamic dispersion and first-order temporal and spatial autoregressions. In the \texttt{glmSTARMA} notation, this corresponds to:
\begin{align*}
\boldsymbol{\phi}_t &= \boldsymbol{\delta} + \sum_{l=0}^1 {\beta}_{1,l} \widetilde{\boldsymbol{W}}^{(l)} h_1(\boldsymbol{Y}_{t-1}), \\
\boldsymbol{\zeta}_t &= \boldsymbol{\widetilde{\delta}} + \sum_{l=0}^1 {\widetilde{\beta}}_{1,l} \widetilde{\boldsymbol{W}}^{(l)} \widetilde{h}_{2}(\boldsymbol{d}_{t-1}),
\end{align*}
where $h_1(\cdot)$ and $\widetilde{h}_2(\cdot)$ are transformations associated with the link functions $g_1(\mu) = \psi$ and $g_2(\phi) = \zeta$, both specified as logarithmic links, and $\{d_t\}$ are conditional deviances. Furthermore, $\boldsymbol{W}^{(\ell)}$ and $\widetilde{\boldsymbol{W}}^{(\ell)}$ are the spatial weight matrices of spatial order $\ell$. For $\ell = 0$, both are the identity matrix $\mathbf{I}_p$, representing temporal self-dependencies. For lag $\ell = 1$, they reduce to the row-normalized adjacency matrix, same as one of the tvd-SPINGARCH fits.

This formulation highlights a structural distinction between the two frameworks regarding their dynamic dispersion mechanisms. While the proposed tvd-SPINGARCH incorporates purely observation-driven INGARCH-type dynamics, \texttt{glmSTARMA}'s dynamic dispersion structure evolves as a function of historical conditional deviances ($\boldsymbol{d}_{t-1}$).

Another aspect that warrants remark is that \texttt{glmSTARMA} assumes spatially uniform (global) parameters. Specifically, the parameters $\beta_{j,l}$ and $\widetilde{\beta}_{j,l}$ are indexed by temporal lag $j$ and spatial lag $l$, but do not vary across individual spatial components (microregions). For instance, $\beta_{j,l}$ represents the universal effect of temporal lag $j$ at spatial lag $l$ across the entire domain, resulting in a parsimonious set of scalar outputs ($\beta_{1,0}, \beta_{1,1}, \widetilde{\beta}_{1,0}, \widetilde{\beta}_{1,1}$) for a first-order spatio-temporal structure. In addition to that, the intercept vectors $\boldsymbol{\delta}$ and $\boldsymbol{\widetilde{\delta}}$ are assumed to be spatially homogeneous by default, such that $\boldsymbol{\delta} = \delta_0 \mathbf{1}_p$ and $\boldsymbol{\widetilde{\delta}} = \widetilde{\delta}_0 \mathbf{1}_p$ (i.e., the homogeneous \texttt{glmSTARMA} specification). Although the software allows for location-specific (inhomogeneous) intercepts, fitting this inhomogeneous variant failed to converge on the TB dataset. Together, these characteristics highlight that \texttt{glmSTARMA} provides flexibility in accommodating temporal and spatial lags, but is less adaptable to location-specific dynamics. 

The empirical performance of the competing model specifications is evaluated next. In addition to benchmarking the tvd-SPINGARCH model against its competitors, an important aim is to evaluate the specific impact of the spatial weight structure. To this end, the tvd-SPINGARCH model is fitted with the discrete adjacency-based matrix (NB tvd-SPINGARCH Adjacency) and the continuous parametric Mat\'{e}rn formulation (NB tvd-SPINGARCH Matérn). Table~\ref{tab:model_comparison} summarizes the resulting maximized log-likelihood, AIC, and BIC values across the fitted models. The NB \texttt{glmSTARMA} yields the worst performance overall, demonstrating that the high parsimony of this global specification comes at the direct expense of empirical fit. In contrast, all remaining specifications (including the Poisson SPINGARCH) estimate microregion-specific parameters. As expected, the NB tvd-SPINGARCH models also provide a substantial improvement over the Poisson benchmark. Finally, the information criteria are essentially non-decisive regarding the spatial weight structure: negligible differences in AIC and BIC are observed between the discrete adjacency and continuous Mat\'{e}rn formulations.

\begin{table}[ht!]
    \centering
    \small
    \renewcommand{\arraystretch}{1.3}
\begin{tabular}{lcccc}
\toprule
\textbf{Model} & \textbf{Spatial operator} & \textbf{Log-likelihood} & \textbf{AIC} & \textbf{BIC} \\
\midrule
NB tvd-SPINGARCH & Adjacency & $-465.31$ &  $937.94$ & $966.39$\\
NB tvd-SPINGARCH & Matérn  & $-$465.21 & 937.25 & 965.77\\
Poisson SPINGARCH & Adjacency & $-$487.91 & 979.48 & 993.71\\
NB glm-STARMA & Adjacency & $-510.53$ & 1021.18 & 1021.65\\
\bottomrule
\end{tabular}
\caption{\it Comparison of likelihood, AIC and BIC values for the NB tvd-SPINGARCH, Poisson SPINGARCH and NB glm-STARMA models across Voronoi and Matérn spatial structures. All values are presented in scale $10^{-2}$.} \label{tab:model_comparison}
\end{table}

\begin{figure}[ht!]
    \centering
    \includegraphics[width=0.95\linewidth]{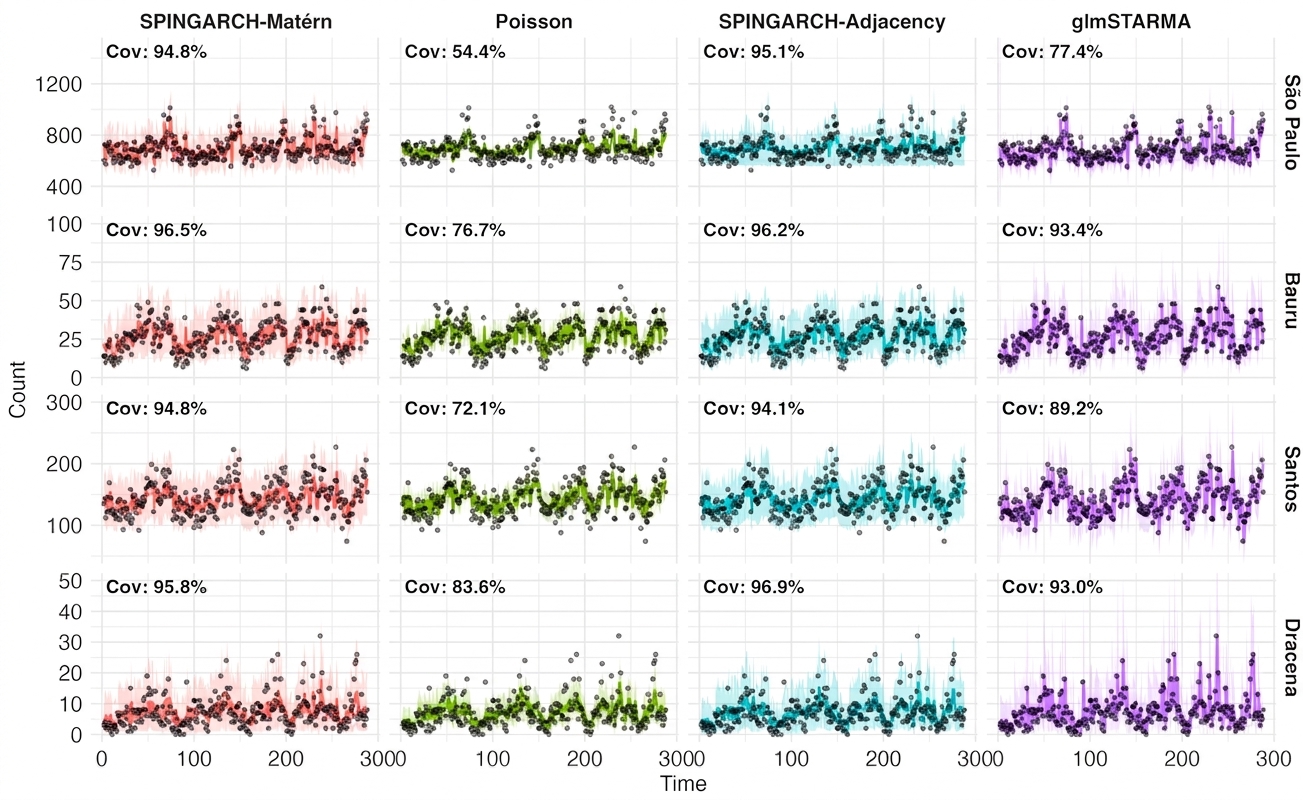}
    \caption{Observed monthly Tuberculosis counts (dots), fitted conditional means (solid lines), and corresponding 95\% confidence intervals (shaded bands) across four representative microregions (rows) for each candidate model specification (columns). The empirical coverage percentage (\texttt{Cov}) of the 95\% confidence intervals is reported in the top-left corner of each panel.}
\label{fig:fitted_microregions}
\end{figure}

To further assess the adequacy of the fitted models, we examine their fit in terms of microregion-specific TB series. The four representative microregions Bauru, Dracena, Santos, and S\~{a}o Paulo are chosen for plotting. Figure~\ref{fig:fitted_microregions} displays the observed counts versus fitted conditional means, where models are arranged as columns and microregions as rows. Observed values are plotted as points, solid lines indicate fitted conditional means, and shaded regions represent 95\% confidence intervals. The empirical coverage percentage reported in the top-left corner of each panel indicates the relative frequency with which confidence intervals contain the observed values. 

The visual comparison highlights striking differences in empirical uncertainty quantification across the competing models. Both proposed tvd-SPINGARCH specifications (Mat\'{e}rn and Adjacency) consistently achieve empirical coverage close to the nominal 95\% level across all four microregions. This accuracy stems from their flexible, location-specific dynamic dispersion mechanism, which allows prediction intervals to naturally expand and contract alongside localized volatility spikes. Conversely, the equidispersed Poisson model generating artificially narrow confidence intervals that result in very low empirical coverage. Finally, \texttt{glmSTARMA} maintains reasonable coverage in lower-volume areas like Dracena (93.0\%) and Bauru (93.4\%). Its performance degrades in larger urban centers like S\~{a}o Paulo (77.4\%) and Santos (89.2\%), reflecting the limitation of its global parameterization in scaling across heterogeneous spatial units.

Empirical coverage clearly favours the NB tvd-SPINGARCH framework, but the analysis remains indecisive regarding the choice of spatial weights. To discriminate between the two, we perform a simulation study to evaluate how well the alternative models reproduce the true underlying spatio-temporal dependencies. To this end, we generate $N$ replicated datasets of length $T$ from each model.

Two metrics are elaborated on top of $\text{Cor}(Y_{i,t}, Y_{j,t-1})$ and $d_{ij}$ to quantify the relationship between first-lag correlations and physical distances. First, an empirical correlation curve is estimated by partitioning the range of inter-region distances into $K$ disjoint intervals $B_k = [c_{k-1}, c_k)$ for $k = 1, \dots, K$, where $0 = c_0 < c_1 < \dots < c_K = \max_{i,j} d_{ij}$. Let $S_k = \{(i,j): i \neq j, d_{ij} \in B_k\}$ denote the set of region pairs whose geographic distance falls within bin $B_k$, and let $|S_k|$ be its cardinality. The binned average lag-1 spatial cross-correlation for interval $B_k$ is 
$$ \bar{\rho}_k = \frac{1}{|S_k|} \sum_{(i,j) \in S_k} \text{Cor}(Y_{i,t}, Y_{j,t-1}). $$
We can then use the discrepancy between the empirical and the replicated curves as a measure of how well the observed spatial behaviour is captured by each model. Formally, we compute for each replication $r$
$$ \text{Curve Error} = \sum_{k=1}^K \left| \bar{\rho}_k^{\text{obs}} - \bar{\rho}_k^{\text{r}} \right|. $$
where $\bar{\rho}_k^{\text{obs}}$ and  $\bar{\rho}_k^{\text{r}}$ are the observed and replication-derived curves. 

The second measure we consider is a distance-weighted error. Let $\mathbf{R}^{(1)} = [\text{Cor}(Y_{i,t}, Y_{j,t-1})]_{i,j=1}^p$ denote the spatial Lag-1 cross-correlation matrix, excluding diagonal elements ($i = j$). For the $r$-th replicate, the distance-weighted error is defined as:
$$ \text{Weighted Error}_r = \frac{1}{p(p-1)} \sum_{i \neq j} d_{ij} \left| \text{Cor}_{\text{obs}}(Y_{i,t}, Y_{j,t-1}) - \text{Cor}_{r}(Y_{i,t}, Y_{j,t-1}) \right|, $$
where subscripts $\text{obs}$ and $r$ denote statistics computed from the observed data and the $r$-th replicate, respectively. By scaling cross-correlation errors linearly with geographic distance $d_{ij}$, this diagnostic heavily penalizes misestimated spillovers across distant microregions. 
This is especially important to evaluate under the Matérn form to ensure that it properly damps dependencies over large distances. 

\begin{figure}[ht!]
    \centering
    \includegraphics[width=0.7\linewidth]{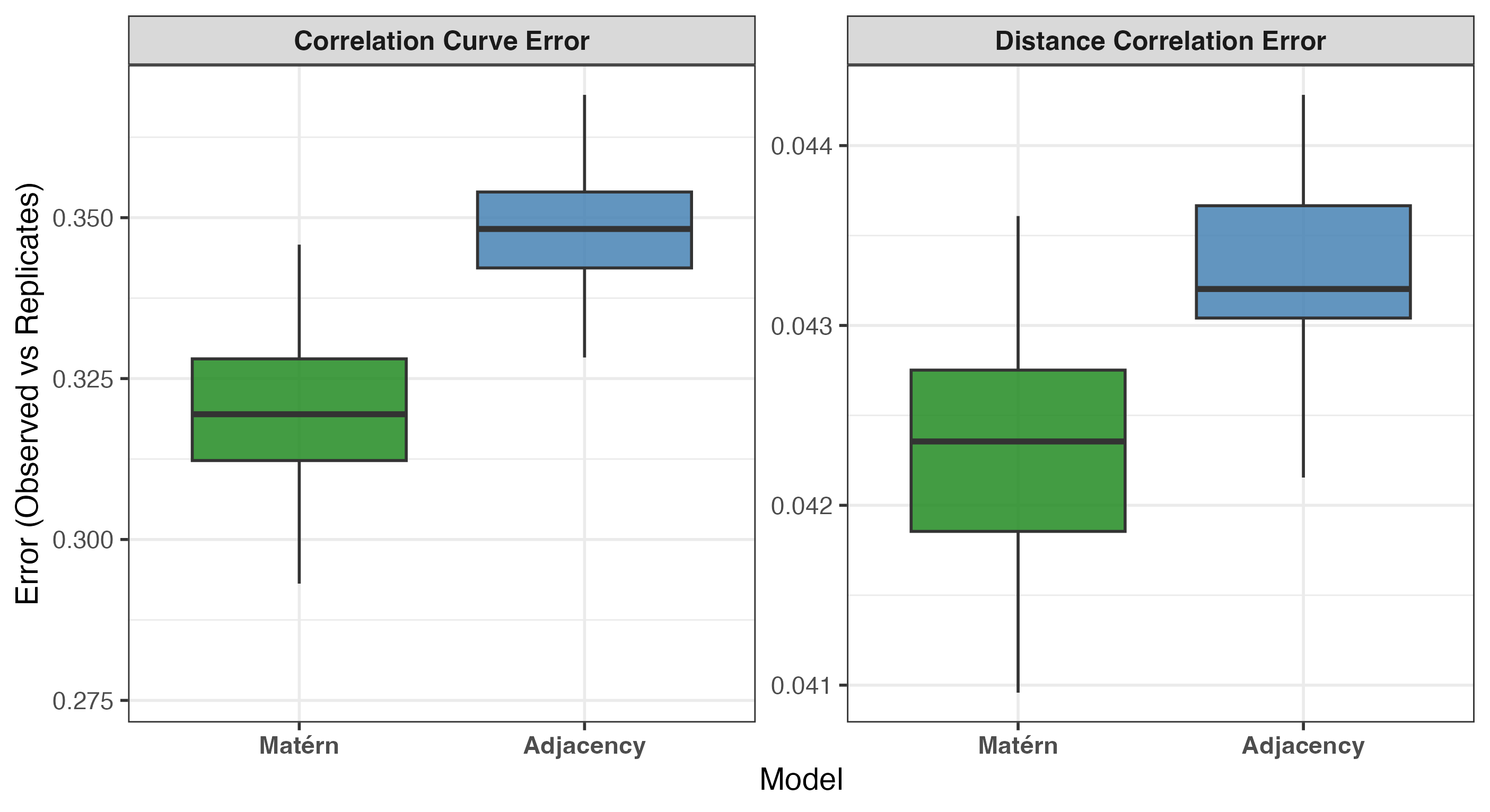}
\caption{Comparison of spatial structural fidelity between Mat\'{e}rn and Adjacency spatial weighting mechanisms across $N = 1000$ data replications. Boxplots display the distribution of errors between observed and simulated statistics for the binned lag-1 correlation decay curve (left panel) and the distance-weighted lag-1 cross-correlation error (right panel).}
\label{fig:spatial_eval}
\end{figure}

Results are displayed in Figure~\ref{fig:spatial_eval}, where boxplots are constructed from $N = 1000$ replications of the study. Both metrics indicate that the Mat\'{e}rn formulation produces a spatial structure that aligns more closely with the observed TB data.  Together, these diagnostics demonstrate that the Mat\'{e}rn model offers a clear advantage in capturing both strong short-range correlations and far-field spatial decay.

\begin{figure}[ht!]
    \centering
    \includegraphics[width=0.95\linewidth]{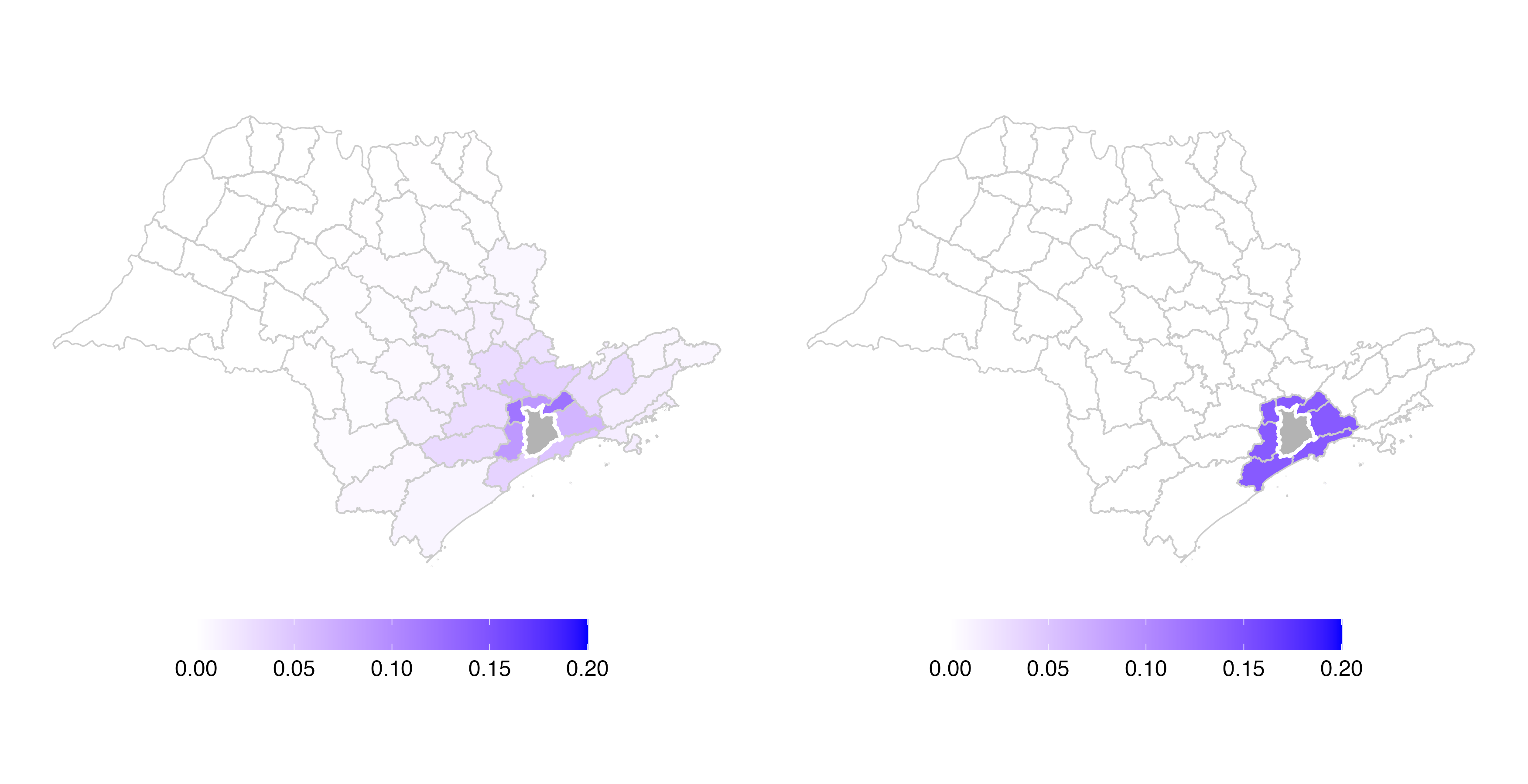}
\caption{Spatial weight distributions for the S\~{a}o Paulo microregion (shaded in grey) under the continuous Mat\'{e}rn (left) and discrete Adjacency (right) specifications.}
\label{fig:weights_sp}
\end{figure}

To illustrate the structural differences between the two specifications, Figure~\ref{fig:weights_sp} visualizes the spatial weight allocations for S\~{a}o Paulo (shaded in grey). In the adjacency scheme (right panel), weight is distributed uniformly across contiguous microregions, with non-adjacent areas receiving zero weight. On the other hand, the Mat\'{e}rn weighting scheme (left panel) incorporates continuous spatial proximity derived from the distances between regional centroids. Consequently, adjacent microregions are not weighted equally, and non-contiguous microregions retain non-zero, continuously decaying weights.

In summary, applying the NB tvd-SPINGARCH model to São Paulo Tuberculosis surveillance highlights that spatiotemporal dispersion is an epidemiological necessity, enabling reliable uncertainty quantification where fixed-dispersion baselines fail. By accurately capturing both volatility surges and spatial distance decay, the proposed framework might be helpful to equip public health authorities with a principled surveillance tool to anticipate localized outbreaks, evaluate regional interventions, and optimize resource allocation.


\section{Concluding remarks}\label{sec:conclusion}

Motivated by the complex spatiotemporal dynamics of Tuberculosis (TB) across São Paulo state, this paper has introduced a negative binomial spatial INGARCH model with dynamic dispersion (NB tvd-SPINGARCH) to advance disease surveillance modeling. By extending the univariate framework of \cite{barretosouzaetal2026} to a multivariate spatial setting, our methodology yields a realistic representation of infectious disease transmission where both the conditional mean and conditional dispersion evolve dynamically over time and space.

Our empirical investigation into monthly TB notifications across 61 microregions over 24 years yields critical insights that might be helpful for epidemiological monitoring and public resource allocation. First, the results demonstrate pronounced spatial heterogeneity, with baseline incidence heavily concentrated in dense metropolitan hubs such as São Paulo city and Santos. Second, the incorporation of dynamic dispersion proved essential: accounting for time-varying dispersion enabled the model to adjust prediction intervals dynamically during periods of heightened volatility, achieving nominal 95\% coverage where fixed-dispersion and Poisson baselines severely underperformed. Third, comparing spatial specifications revealed that inter-regional transmission is predominantly short-ranged, with both boundary adjacency and Matérn distance-decay structures effectively capturing local spatial spillovers.

From a statistical perspective, the proposed framework provides a rigorous foundation for spatial count modeling. Combining conditional maximum likelihood with a two-step profile-likelihood strategy renders joint parameter estimation computationally feasible across multi-region networks, while parametric bootstrap inference offers reliable uncertainty quantification. Furthermore, embedding the Matérn correlation family into the spatial weight structure provides a flexible, data-driven alternative to rigid administrative boundaries, allowing spatial smoothness and scale to be estimated directly from surveillance data.

Beyond TB surveillance in São Paulo, we believe that the methodology developed here offers a versatile tool for analyzing high-dimensional, volatile count processes in epidemiology and environmental monitoring. 

\paragraph{Software and data availability.}
The \texttt{R} implementation of the NB tvd-SPINGARCH model can be obtained from the authors upon request. The TB data are publicly available and can be obtained from the Department of Information and Informatics of the Brazilian
Unified Health System (DATASUS).

\paragraph{Conflict of interest.} None to be declared.


\end{document}